\documentclass[preprint,showpacs,amsmath,amssymb]{revtex4}
\usepackage{graphicx}
\usepackage{psfrag}
\usepackage{epsfig}
\usepackage{rotating}

\begin{document}

\title{Geographical dispersal of mobile communication networks}
\author{Renaud Lambiotte$^{1,2}$, Vincent D. Blondel$^1$, Cristobald de Kerchove$^1$, Etienne Huens$^1$,   Christophe Prieur$^3$, Zbigniew Smoreda$^3$, Paul Van Dooren$^1$}

\affiliation{
$^1$Department of Mathematical Engineering, Universit\'e catholique de Louvain,  4 avenue Georges Lemaitre, B-1348 Louvain-la-Neuve, Belgium \\
$^2$ Institute for Mathematical Sciences, Imperial College London, 53 Prince's Gate, South Kensington campus, SW72PG, UK\\
$^3$Sociology and Economics of Networks and Services Department, Orange Labs, F-92794 Issy-les-Moulineaux, France
}

\date{\today}

\begin{abstract}
In this paper, we analyze statistical properties of a communication network constructed from the records of a mobile phone company. The network consists of 2.5 million customers that have placed 810 millions of communications (phone calls and text messages) over a period of 6 months and for whom we have geographical home localization information. It is shown that the degree distribution in this network has a power-law degree distribution $k^{-5}$ and that the probability that two customers are connected by a link follows a gravity model, i.e. decreases like $d^{-2}$, where $d$ is the distance between the customers. We also consider the geographical extension of communication triangles and we show that communication triangles are not only composed of geographically adjacent nodes but that they may extend over large distances. This last property is not captured by the existing models of geographical networks and in a last section we propose a new model that reproduces the observed property. Our model, which is based on the migration and on the local adaptation of agents, is then studied analytically and the resulting predictions are confirmed by computer simulations.
\end{abstract}

\maketitle

\section{Introduction}

In recent years, complex network science has been a very active inter-disciplinary research area. The empirical analysis of a large number of technological, information and social networks \cite{revN,revB} has revealed that  alike networks often share common topological properties. For instance, it is now well-known that social networks exhibit the small-world property and assortative correlations between the degrees of neighboring nodes \cite{newman0}. From a theoretical point of view, many of these universal properties are now well understood and simple theoretical models have been shown to reproduce quite well the empirical evidence, e.g. scale-free degree distributions emerge in growing network due to preferential attachment mechanisms \cite{bara}. The models also allow to understand the role played by the network topology on the spreading of information  \cite{Watts02,boguna,sood,lam0,centola}.

Complex networks are usually assumed to be homogeneous, i.e. the nodes are a priori equivalent and the only way to differentiate them is to compare their topological properties. This approach is sometimes conceptual, as one of the goals of complex network theory is to deduce the function or state of a node only from its location in the network, but it is also much more pragmatic, as it is usually difficult to find reliable information about the ``internal properties" of a node, such as the taste or opinion of an individual. The geographical position of the nodes is, though, such a relevant and unambiguous property. Indeed, the nodes of a network may have positions in space and, in many cases, it is reasonable to assume that geographical proximity plays a role in deciding how to connect the nodes \cite{bart,kitch,kleinberg,wong,bogu}. Restriction of long-range links has been observed in many real networks \cite{hayashi}, such as the Internet  \cite{internet}, road networks and flight connections  \cite{airline} and brain functional networks \cite{eguiluz}.  The length statistics of the links may vary from one network to another. For instance, the road network has only very short links, due to obvious physical constraints, while ``higher dimensional" networks such as the Internet and airline networks have much broader length distributions. In the case of socio-economic networks, though, a power-law decrease $\sim 1/d^\alpha$ of the flux or interaction between two places seems to be universal \cite{gravity0} and it has been observed in many situations such as the International Trade Market \cite{gravity1} or traffic flows between cities \cite{gravity2}. In most of these socio-economic networks the exponent $\alpha$ is  very close to 2, which suggests to name these systems gravitational \cite{pr1,pr2}, as a metaphor of physical gravity as described in Newton's law of gravity. In that case, the size of the geographical entity, e.g. the number of inhabitants, plays the role of its mass.

In this paper, we consider the geographical dispersal of a social network. We do not restrict the scope to the length of the links, as in these previous studies, but generalize instead our analysis to the dispersal of triangular motifs. To do so, we construct a network where the nodes are the customers of a mobile phone company and the links represent calls between these customers. Let us stress that it is increasingly popular to study mobile phone data in order to explore large-scale social systems and to reveal how individuals interact with each other \cite{onnela,onnela2,candia,hidalgo}. It seems obvious that in such a network the geographical location of the individuals is an important communication factor \cite{z0}. This is due to the fact that people are more likely to form social ties with others who live close by, so that mobile phone communication ought to reflect this underlying dependence of social bonding on distance. Our main objective in this contribution is to characterize the geographical properties of the communication network, which is of crucial importance if one wants to predict how ideas and information spread geographically. As a first step, we focus on the geographical length of the links and show that the probability that two people call each other follows a gravity model. Then, we generalize our geographical analysis to communication triangles which are the most basic measure of the presence of communities and of the cohesion of the social system.
Triangles are well-known to be numerous in social systems due to the transitivity of social interactions \cite{watts} and to social balance \cite{triads}.
It is shown here that the probability for a link to belong to a triangle goes to a constant when its length is sufficiently large, a feature that is not observed in classical network models \cite{watts,wattsb} exhibiting the small-world property, i.e. a high clustering coefficient and a short diameter. We then propose a simple model in which agents migrate and adapt to their local environment and that produces the observed property. For our model, we derive analytically the geographical extension of triangles in the system.

\section{Data analysis}

\subsection{Data description}

\begin{figure}[]
\includegraphics[angle=-90,width=0.55\textwidth]{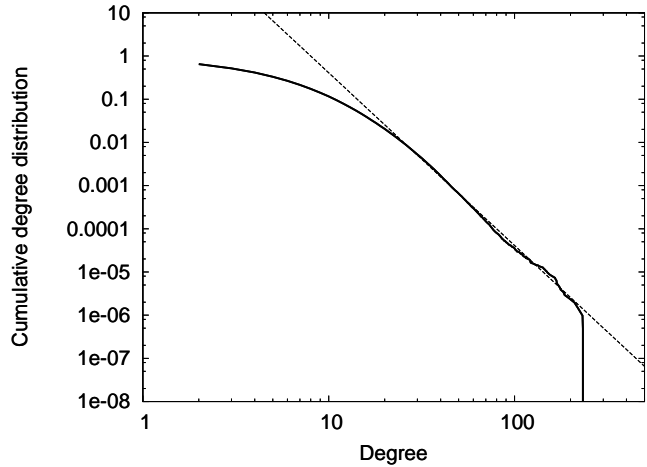}
\caption{Cumulative degree distribution $\sum_{k=K}^{\infty} p_k$ as a function of the degree $K$. The cumulative distribution is very well fitted by the power-law $K^{-4}$ (dashed line), which implies that the tail of the degree distribution $p_k$ behaves like $k^{-5}$ \cite{pl1,pl2}.}
\label{distrib}
\end{figure}

The data that we consider consist of the communications made by over 3.3 million customers from a Belgian mobile phone company over a period of 6 months.  Each customer is identified by a surrogate key to which several entries are associated, such as his age, his sex, his language and the zip code of the location to which the bill is sent. Several profiles are lacking in the database and we therefore restrict the scope to the 2.5 million customers whose profile is complete. Moreover, for the sake of simplicity, we focus only on mobile phone calls and text messages and discard other types of communications, such as voice mail and data calls. From now on, phone calls and text messages will be termed ``calls" indifferently.
In order to construct the communication network, we have also filtered out calls involving other operators (there are three main operators in Belgium), incoming or outgoing, and have kept only those transactions in which both the calling and receiving individuals are clients of the same operator. By doing so, we keep 810 millions calls between the 2.5 million customers. The resulting network is composed of 2.5 million nodes and of 38 million links which are weighted (the weight may be the number of phone calls or the total communication time) and directed (from the outgoing to the incoming).  However, many of these interactions are only one-way, which suggests that they correspond to single events and that the two interacting individuals do not actually know each other. In order to eliminate these ``accidental calls", we have kept links between two individuals $i$ and $j$ only if there had been at least six reciprocated pairs of calls between them during the 6 months time interval. It turns out that the exact number of reciprocated calls is not essential; all qualitative observations described in this paper have been shown to remain true for the same network filtered on four or eight reciprocated phone calls.   The resulting undirected, unweighted network is composed of $5.4 $ million links and has therefore an average degree of $4.3$. It exhibits typical properties of social networks, such as a broad degree distribution $p_k$ whose tail is very well-fitted by the power-law $k^{-\gamma}$, with $\gamma=5$ (see Fig.~\ref{distrib}). This relatively high value should be compared to the values $\gamma=8.4$ and $\gamma=2.1$  observed for another mobile phone network \cite{onnela2,biely} and for landlines \cite{doro}.

\subsection{Gravity model}

\begin{figure}[]
\includegraphics[angle=-90,width=0.55\textwidth]{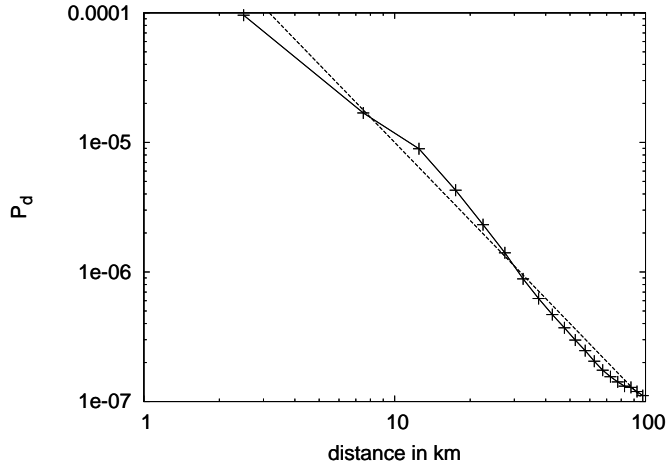}
\caption{We plot the probability $P_d$ that two people living at a distance $d$ are connected by a link in a log-log scale. The dashed line is the power-law $d^{-2}$.}
\label{links}
\end{figure}

Let us now focus on the spatial separation of connected nodes of the mobile phone network. In order to approximate these geographical distances we used the zip code provided by the customer for billing purposes; there are 1145 different zip codes in the database and the distance between two zip code areas was calculated by using their geographical coordinates. This method does not allow to evaluate distances between people living in the same zip code area and the distance is assumed to be zero in that case. It is useful to define the number $L_d$ of links of length $d$, i.e. the number of pairs of connected nodes separated by a distance $d$,  and the total number $N_d$ of pairs of people (connected or not) who are separated by a distance $d$. The probability that two individuals separated by a distance $d$ are related by a link is therefore $P_d=L_d/N_d$. Practically, we have looked at the length of the links with a resolution of 5 km, which is a typical size for the distance between two zip code areas. The empirical analysis shows that $P_d$ is very well approximated by a gravity model $\sim d^{-2}$, over a large range of distances (see Fig.~\ref{links}). 

\begin{figure}[]
\includegraphics[width=0.5\textwidth]{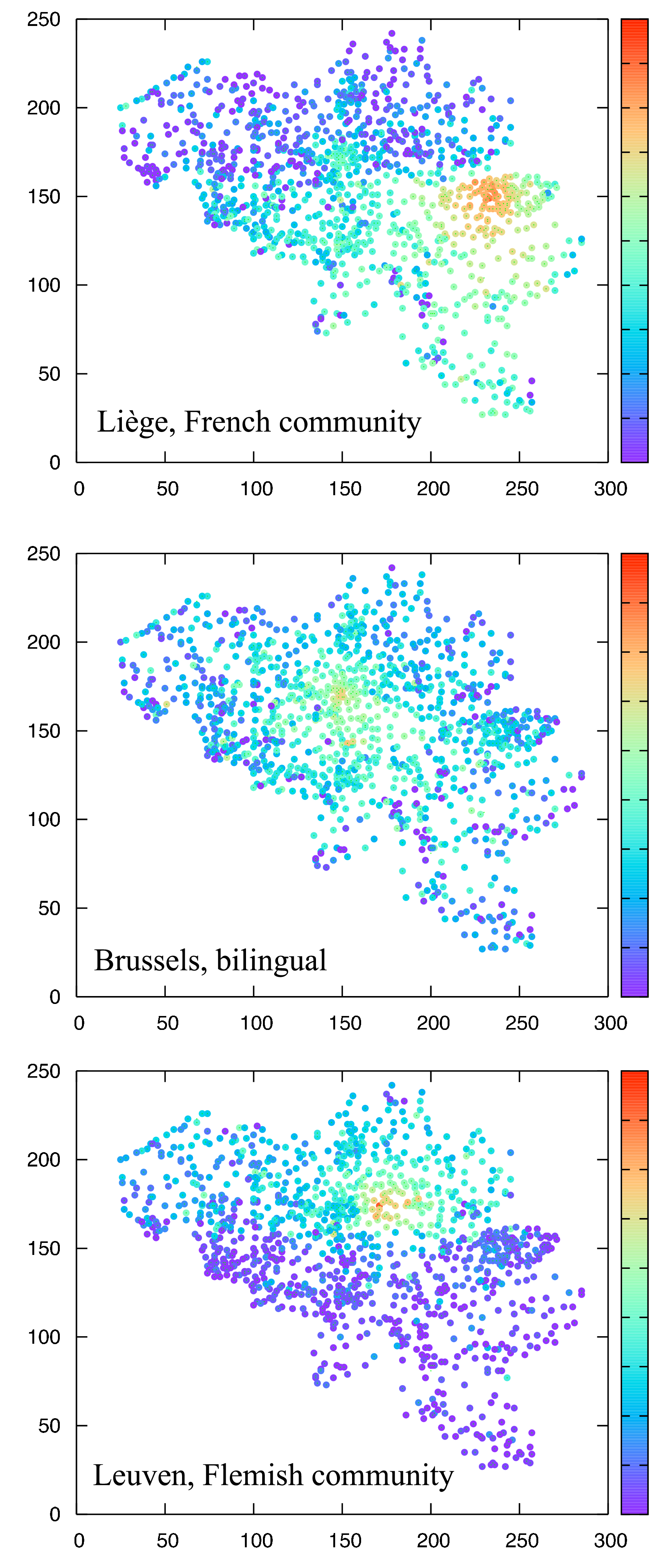}
\caption{Color representation of the probability that a customer from Li\`ege, Brussels and Leuven respectively has a link with a customer from another zip code area in Belgium. The more red (purple) a point is, the higher (lower) is the probability to have a link to that zip code area. One observes a clear north-south asymmetry, due to the different languages spoken in the communities. Of the three cities, only Brussels seems to call indifferently to the north and to the south.}
\label{belgium}
\end{figure}

The above analysis implicitly assumes that the system is homogeneous and isotropic, conditions that are far from obvious in a realistic environment. In that sense, the case of Belgium is exceptional in that the two main language communities live in different regions of Belgium (roughly speaking, the south is French speaking and the north Flemish speaking). This geographical segregation  leads to a strong north-south asymmetry (see Fig.~\ref{belgium}) in the distribution of the calls.

\begin{figure}[]
\includegraphics[angle=-90,width=0.55\textwidth]{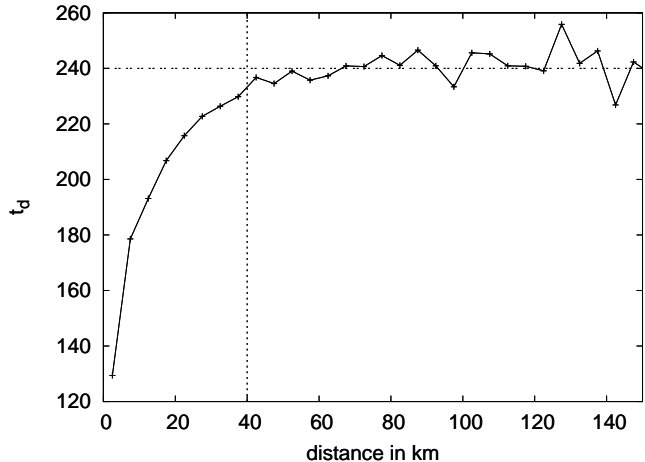}
\caption{We plot the average duration  $t_d$ of a phone call (in sec) as a function of the distance. This duration $t_d$, which is evaluated over the whole 6 month period, increases until it reaches the plateau value of $240$ sec. The average of $t_d$ over all phone calls is $<t_d>=157$ sec.}
\label{timeAlone}
\end{figure}

It is also interesting to note that the average duration of phone calls increases with the distance but reaches a plateau around $d=40$ km (see Fig.~\ref{timeAlone}). The increase at the communication time with the distance has already been observed in previous studies of residential-fixed phone usage \cite{shortcalls0,shortcalls0b}. This is due to the fact that people, when they live at short distances, frequently meet and communicate face-to-face. Phone calls are therefore short and functional  (``Let's meet at 8PM at the pub"), and aim at the coordination and synchronization of the individuals' activities. In contrast, at longer distances the telephone is one of the main communication medium and is a crucial resource in order to maintain a relationship. In that case, people ask about each other and take the time to talk  \cite{shortcalls1}. The plateau beyond $40$ km suggests that distance ceases to be a relevant parameter once the two interlocutors are far enough from each other.

\begin{figure}[]
\includegraphics[angle=-90,width=0.55\textwidth]{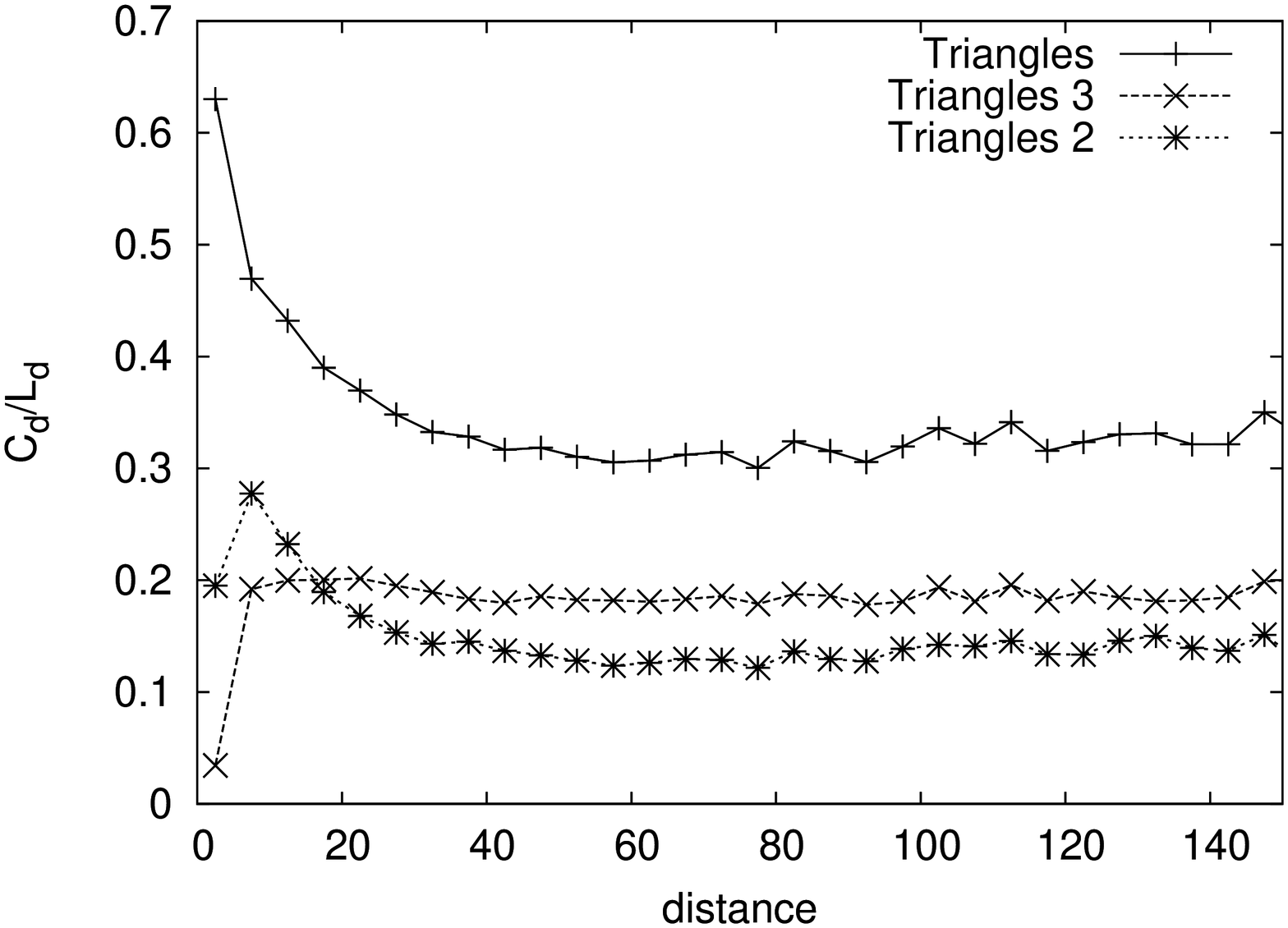}
\caption{Probabilities $c_d$, $c_d^2$ and $c_d^3$ that a link of length $d$ belongs to a communication triangle, to a triangle of type $2$ ($i.e.,$ extended over two different zip code areas) and to a a triangle of type $3$ ($i.e.,$ extended over three different zip code areas) respectively. The quantity $c_d$ is seen to decrease until it reaches the plateau value $0.32$.}
\label{triangles}
\end{figure}

\subsection{Communication triangles}

Let us now focus on the geographical dispersal of more complicated motifs \cite{motifs}. To our knowledge, the only work going in that direction is that of \cite{palla} where the authors show that the zip-codes of mobile phone users inside a community are highly correlated, thereby indicating that people inside the same community have a tendency to contain people living in the same neighborhood. In \cite{palla}, however, this analysis was performed in order to check the validity of the community detection method and not as an objective per se.
In this paper, we focus instead on three-cliques (triangles) which are the first generalization of two-cliques (links). This choice is partly motivated by the fact that triangles are typical motifs of social networks, due to the transitivity of friendship relations.

In order to evaluate the spatial extension of the $T=1\, 840\, 552$ triangles found in the network, we have measured the number $C_d$  of links which have a length $d$ and which belong to a triangle. By construction, the quantity $c_d=C_d/L_d$ is therefore the probability that a link of length $d$ belongs to a triangle. One observes that $c_d$ decreases with the distance, thereby showing that shorter links have a higher probability to belong to triangles than longer links (see Fig.~\ref{triangles}). However, $c_d$ ceases to decrease at around $40$ km, where it reaches a constant value around $0.32$. Beyond this value, links belonging to triangles  have the same spatial statistics as any other link in the system.
Interestingly, the crossover takes place at approximately the same distance $40$ km as the average communication time does. This suggests the existence of two regimes of communication: a short-distance ``face-to-face" regime characterized by short communications and a high clustering coefficient, and a long-distance regime characterized by longer communications and a smaller clustering coefficient. It is important to note, however, that $c_d$ remains quite large in the second regime and that the total variation of $c_d$ remains limited in the distance interval, i.e. $c_d$ decreases of only $50 \%$ from $d=0$ to $d=150$.

Finally, let us stress that links may have different statistics depending on the different kind of triangles to which they belong. To show so, the triangles are divided in three classes depending on the number of different zip codes present in the triangle. In the first class, all individuals in the triangle have the same zip code. There are $T_1=703\, 137$ such triangles. In the second class, individuals live in two different zip code areas, which takes place for $T_2=726\, 076$ triangles. The third class consists of the $T_3=411\, 339$ triangles whose individuals live in three different zip code areas. Then, we have measured the number $C_d^i$ of links which have a length $d$ and which belong to a triangle of class $i$. By construction, the behavior of $C_d^1 \sim \delta_{d0}$ is trivial because all the nodes have the same zip code.
The probability $c_d^2=C_d^2/L_d$ is seen to exhibit a decrease at short distances, followed by a plateau, while $c_d^3=C_d^3/L_d$ is constant over the whole distance interval, thereby showing that the probability of links to belong to such triangles does not depend on the distance.

\begin{figure}[]
\includegraphics[width=0.9\textwidth]{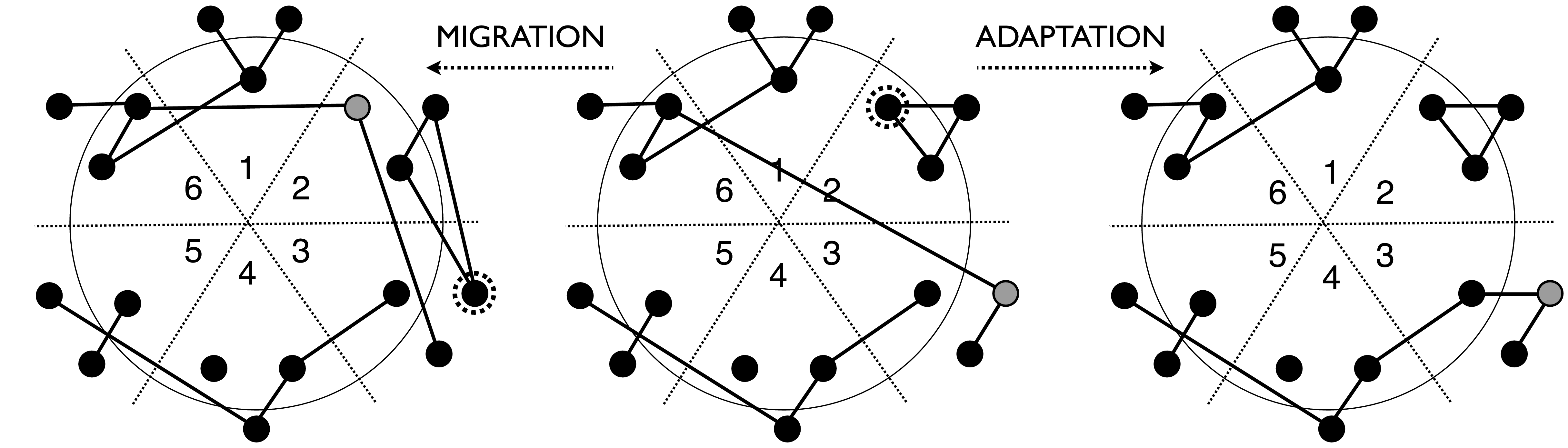}
\caption{Illustration of an update of the network dynamics. The system is composed of $S=6$ sites and $N=18$ nodes. At one time step, the grey node is randomly selected. With probability $p$, the selected node migrates by switching its position with another randomly selected node (surrounded). With probability $1-p$, the selected node breaks its old links and creates new links with the other nodes of his site.}
\label{model}
\end{figure}

\section{Model of geographical network}

\subsection{Description}

The fact that $c_d$ is almost constant over the whole distance interval and, especially the fact that $c_d$ does not go to zero at large distances, deserves an attentive look. Indeed, such a behavior is not reproduced by models of geographical networks \cite{kleinberg,hayashi} where some of the links of a regular lattice are randomly redistributed through the system. By construction, such networks may be composed of many triangles, but mainly local triangles, i.e. triangles composed of neighboring nodes. 
A similar behaviour also takes place in the Watts-Strogatz model for small-world networks \cite{watts}, which can be viewed as a one-dimensional geographical network where the distance is the number of hops between sites on the underlying lattice.

In order to reproduce a network with a non-vanishing number of extended triangles, we propose instead to consider a system where agents move geographically and keep their links after they have moved. Let us consider a prototypal model where agents move on a periodic one-dimensional lattice of length $S$; see Fig. \ref{model}. We assume that there are three agents at each site $i$ so that the total number of agents in the system is $N=3S$. At each time step, one agent is randomly selected. With probability $p$, the selected agent moves in the system and carries its links. For the sake of simplicity, we assume that an agent at site $i$ may attain any site $j$, independently of the distance between $i$ and $j$. Moreover, in order to ensure that each site is composed of exactly three agents at every step, we also select an agent of the site $j$ and move it to $i$ so that in practice the two agents simply exchange their positions, but keep unchanged the agents to which they are connected. With probability $1-p$, in contrast, the selected agent breaks its current links and creates new links with its two neighbors. Consequently, the dynamics is driven by the competition between the migration of the agents, which extends and deforms the triangles inside the system, and the adaptation of the agents to their local environment, which replaces long-distance links by short-distance links and therefore favors the creation of local triangles.

Before going further, one should stress that the mean-field assumption for the length of the jumps simplifies  the analysis significantly, as links are either intra-site ($d=0$) or inter-sites ($d>0$) and the precise value of the distance $d>0$ between two sites is not relevant for characterizing the dynamics. By doing so, one therefore decomposes the system into two levels: the local level where people meet each other and interact, and the distant level of people living far from each other.
Let us also note that the ingredients of our model are reminiscent of the model of \cite{onnelaPRL}, where the system is driven by a competition between cyclic closure and focal closure but also of the model of mobile agents of \cite{pedro} which is based on the motion of particles that create temporary links with the particles that they encountered in the past.

\subsection{Some results}

\begin{figure}[]
\includegraphics[angle=-90,width=0.55\textwidth]{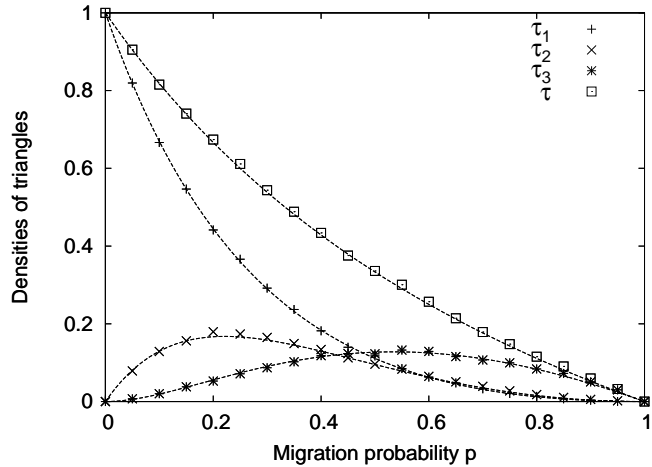}
\caption{Densities of triangles $\tau_1$, $\tau_2$, $\tau_3$ and $\tau$ as a function of the migration probability. The points correspond to computer simulation results and the dashed lines to the theoretical predictions (\ref{s1}), (\ref{s2}) and (\ref{s3}).}
\label{simulation}
\end{figure}

Let us now focus on the statistical properties of the model. As a first step, we focus on the numbers $L_{in}$ and $L_{out}$ of intra-site and inter-sites links. By construction, $L_{in}+L_{out}=L$, where $L$ is the total number of links. It is straightforward to show that $L$ asymptotically goes to $L=N$, because migration does not alter the number of links while adaptation makes the average degree of the node equal to 2. In order to derive the probability $l_{in}\equiv L_{in}/L$, it is useful to take a continuous time limit and to write the rate equation for $l_{in}$

\begin{eqnarray}
\label{}
\partial_t l_{in} = - 4 p ~ l_{in} + 2 (1-p) ~(1-l_{in}),
\end{eqnarray}
The loss term accounts for situations where a node migrates and brings its links to its new site. Such an event occurs with a probability $p$ and $4 ~l_{in}$ are involved in that case, because two nodes move and the average degree of the nodes is $2 ~l_{in}$. The gain term accounts for situations where intra-site links are created by adaptation. We have implicitly assumed that the system is infinitely large in order to evaluate these probabilities. It is straightforward to show that $l_{in}$ asymptotically goes to
\begin{eqnarray}
\label{pred}
l_{in} =\frac{1-p}{1+p},
\end{eqnarray}
which implies that $l_{in}$ goes to 1 and 0, in the limits $p \rightarrow 0$ and $p \rightarrow 1$ respectively, as expected. Let us stress that (\ref{pred}) can also be obtained in a probabilistic way by looking at an arbitrary link $\ell$ and by noting that the expected value of $l_{in}$ is simply the probability that the two nodes connected by $\ell$ belong to the same site. That amounts to consider the probability that the last operation implying one of these $2$ nodes was an adaptation, so that $l_{in}=(1-p)/Z$. The normalization $Z$ comes from the fact that two nodes exchange position in case of migration, so that $Z=(1-p) +  2 p=1+p$.

Let us now focus on the numbers $T_i$ of triangles that extend
over $i$ different sites and on their respective nodes of class
$i$. In order to find the asymptotic values of $T_1$, $T_2$ and
$T_3$, let us write the expected values for $\tau_i\equiv T_i/S$
when $S \rightarrow \infty$. Since $\tau_i$ corresponds to the
probability that a random node $k$ belongs to class $i$, $\tau_1$
is obtained by considering the probability that the last operation
of at least $2$ nodes in the site of node $k$ was an adaptation and thus
\begin{eqnarray}
\label{s1} \tau_1 &=& (\frac{1-p}{1+p})^2.
\end{eqnarray}
$\tau_2$ is obtained by considering the probability that node $k$
was in class $1$ and then it has migrated, or it was in class $2$
with the two other nodes of the triangle in another site and then
it has migrated. Hence,

\begin{eqnarray}
\label{s2} \tau_2 &=& \frac{2p}{1+p}\tau_1 + \frac{1}{3}\cdot
\frac{2p}{1+p}\tau_2 = \frac{2p}{(1+p/3)}\cdot
\frac{(1-p)^2}{(1+p)^2}.
\end{eqnarray}
$\tau_3$ is obtained similarly, 
\begin{eqnarray}
\label{s3} \tau_3 &=& \frac{2}{3}\cdot\frac{2p}{1+p}\tau_2 +
\frac{2p}{1+p}\tau_3 =  \frac{8 p^2}{3+p}\cdot
\frac{(1-p)}{(1+p)^2},
\end{eqnarray}
by considering the probability that node $k$
was in class $2$ with another node of the triangle in the same
site and then it has migrated, or it was in class $3$ and then it
has migrated.

By summing the contributions (\ref{s1}), (\ref{s2}) and
(\ref{s3}), one finds the total number of triangles
$T=T_1+T_2+T_3= S (1-p)/(1+p) $. These theoretical predictions
have been verified by performing computer simulations of the
model. To do so, one considers a system composed of $S=100$ sites
and starts the simulations from a random initial condition. The
number of different triangles is measured after long times, i.e.
$100$ steps/node. The results, that are averaged over $100$
realizations of the process, are in excellent agreement with the
predictions (see Fig.~\ref{simulation}). It is interesting to note
that the numbers of triangles $T$ and $T_1$ are maximum for $p=0$,
as expected, while the numbers of decentralized triangles $T_2$
and $T_3$ are maximum for intermediate values of $p$.

\begin{figure}[]
\includegraphics[angle=-90,width=0.55\textwidth]{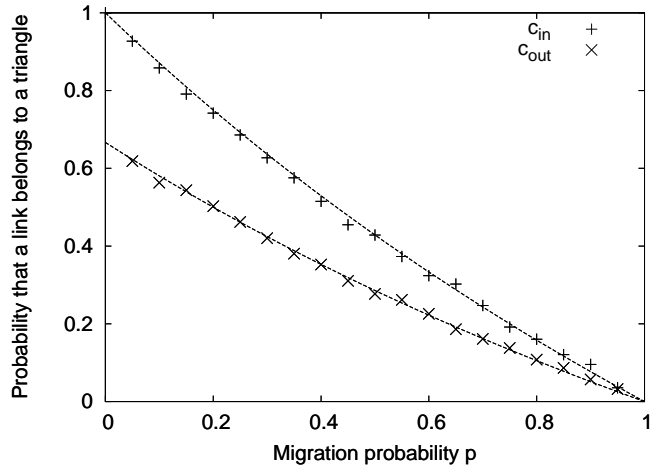}
\caption{Probabilities $c_{in}$ and $c_{out}$ that an intra-site and an inter-sites link respectively belong to a triangle, as a function of the migration probability. The points correspond to computer simulation results and the dashed lines to the theoretical predictions (\ref{final}).}
\label{simulation2}
\end{figure}

Finally, let us stress that it is now straightforward to derive the probabilities $c_{in}$ and $c_{out}$ that an intra-site and an inter-sites link respectively belong to a triangle. Indeed, the total numbers of intra-site and inter-sites links are simply
\begin{eqnarray}
C_{in} &=& 3 T_{1} + T_2 =   \frac{9 (1-p)^2 S}{(1+p) (3+p)}\cr
C_{out} &=& 2 T_{2} + 3 T_3=  \frac{12 p (1-p) S}{(1+p) (3+p)}.
\end{eqnarray}
By combining this result with the prediction (\ref{pred}), the probabilities  $c_{in}$ and $c_{out}$ are found to be
\begin{eqnarray}
\label{final}
c_{in} &=&  \frac{3 (1-p) }{(3+p)}\cr
c_{out} &=&  \frac{2 (1-p) }{(3+p)}.
\end{eqnarray}
These predictions, that have been successfully verified by computer simulations (see Fig.~\ref{simulation2}), confirm that $c_{out}$ does not vanish in the limit $S \rightarrow \infty$, thereby reproducing qualitatively the empirical results exposed in the previous section. Indeed,  $c_{in}$ and $c_{out}$ may be viewed as a coarse-grained version of the probabilities studied in Fig.\ref{triangles}, i.e. $c_{in}$ and $c_{out}$ correspond to $c_d$ for short and long distances respectively, and the fact that $c_{out}$ remains finite therefore implies that the probability to belong to a triangle does not vanish at long distances.

\section{Conclusion}

We have analyzed a large social network where nodes are customers of a Belgian mobile phone company and links correspond to reciprocated phone calls that we identify as their social interactions. We have focused on the geographical component of this social network by first studying the statistics of the length of the links. It is shown that these lengths follow a gravity model, namely the probability that two individuals are connected is inversely proportional to the square of the distance between them. 
It is interesting to note that such networks are known to minimize the delivery time of messages by decentralized algorithms, i.e. algorithms in which individuals only know the locations of their direct acquaintances \cite{kleinberg}. This suggests that the mobile phone network has an optimal topology in order to deliver information through the system.

This analysis is generalized by studying the geographical extension of communication triangles, which are well-known to be characteristic motifs of social networks. Our main result is that the system is composed of many extended triangles, much more than in typical models of geographical networks. We therefore propose a model for the evolution of the social network in order to explain this property. To do so, we focus on a system where agents may either migrate, while carrying their links during their motion and therefore deforming the triangle to which they belong, or adapt to their local environment by breaking their previous links and creating links with their geographical neighbors. By doing so, one couples the motion of the agents to the topology of the social network in a very generic and intuitive way \cite{pedro}, thereby suggesting that more realistic models will behave qualitatively in the same way, $e.g.$ models including a more complicated distribution of migration lengths or preferential attachment processes. The simplicity of our approach has the advantage to allow an analytical treatment and to clarify how the competition between migration and adaptation may influence the geographical dispersal of network motifs.

\acknowledgments This paper presents research results obtained in part by the research group ``Large
Graphs and Networks'' at UCL. The group is funded in part by the Communaut\'e Fran\c caise de Belgique through an ARC and by the Belgian Network DYSCO (Dynamical Systems, Control, and Optimization), funded by the Interuniversity Attraction Poles Programme, initiated by the Belgian State, Science Policy Office. This work has also been supported by the Orange Labs R\&D Research Grant 46143202. The scientific responsibility rests with its authors.

\end{document}